\documentclass[conference]{IEEEtran}
\IEEEoverridecommandlockouts
\usepackage{cite}
\usepackage{amsmath,amssymb,amsfonts}
\usepackage{algorithmic}
\usepackage{graphicx}
\usepackage{textcomp}
\usepackage{xcolor}
\usepackage{booktabs}
\usepackage{array}
\usepackage{tcolorbox}
\usepackage{vtable}
\def\BibTeX{{\rm B\kern-.05em{\sc i\kern-.025em b}\kern-.08em
    T\kern-.1667em\lower.7ex\hbox{E}\kern-.125emX}}

\makeatletter
\newcommand{\linebreakand}{
  \end{@IEEEauthorhalign}
  \hfill\mbox{}\par
  \mbox{}\hfill\begin{@IEEEauthorhalign}
}
\makeatother
\begin{document}

\title{Metadata-based Data Exploration with Retrieval-Augmented Generation for Large Language Models}

\author{
\IEEEauthorblockN{Teruaki Hayashi}
\IEEEauthorblockA{\textit{Department of Systems Innovation, School of Engineering} \\
\textit{The University of Tokyo}\\
Tokyo, Japan \\
hayashi@sys.t.u-tokyo.ac.jp}

\and
\IEEEauthorblockN{Hiroki Sakaji}
\IEEEauthorblockA{\textit{Faculty of Information Science and Technology} \\
\textit{Hokkaido University}\\
Hokkaido, Japan \\
sakaji@ist.hokudai.ac.jp}

\linebreakand

\IEEEauthorblockN{Jiayi Dai}
\IEEEauthorblockA{\textit{Alberta Machine Intelligence Institute} \\
\textit{University of Alberta}\\
Edmonton, Canada \\
dai1@ualberta.ca}

\and
\IEEEauthorblockN{Randy Goebel}
\IEEEauthorblockA{\textit{Alberta Machine Intelligence Institute} \\
\textit{University of Alberta}\\
Edmonton, Canada \\
rgoebel@ualberta.ca}
}

\maketitle

\begin{abstract}
Developing the capacity to effectively search for requisite datasets is an urgent requirement to assist data users in identifying relevant datasets considering the very limited available metadata. For this challenge, the utilization of third-party data is emerging as a valuable source for improvement. Our research introduces a new architecture for data exploration which employs a form of Retrieval-Augmented Generation (RAG) to enhance metadata-based data discovery. The system integrates large language models (LLMs) with external vector databases to identify semantic relationships among diverse types of datasets. The proposed framework offers a new method for evaluating semantic similarity among heterogeneous data sources and for improving data exploration. Our study includes experimental results on four critical tasks: 1) recommending similar datasets, 2) suggesting combinable datasets, 3) estimating tags, and 4) predicting variables. Our results demonstrate that RAG can enhance the selection of relevant datasets, particularly from different categories, when compared to conventional metadata approaches. However, performance varied across tasks and models, which confirms the significance of selecting appropriate techniques based on specific use cases. The findings suggest that this approach holds promise for addressing challenges in data exploration and discovery, although further refinement is necessary for estimation tasks.
\end{abstract}


\section{Introduction}

Despite growing interest in leveraging data sharing and collaboration for value creation, various obstacles hinder the use of data across different disciplines. A key challenge is locating data that aligns with specific interests. Currently, dataset retrieval often relies on basic query matching with metadata titles and dataset contents \cite{chapman2020, miller2018opendatatransparent}. However, due to the highly specialized terminology found in both datasets and metadata, traditional search methods like query matching typically fall significantly short for users lacking domain expertise. These individuals and groups struggle to articulate their interests using specialized terms, making it difficult for them to find the datasets of best value. 

This issue further extends to organizations whose mandate is to provide data. because the generation of metadata necessitates specialized knowledge of the specific data and field. To do so requires substantial effort and resources. Furthermore, data specifically created for third-party use is uncommon, and data providers often lack sufficient incentives to maintain metadata. Consequently, publicly available open data frequently had inadequate metadata \cite{zezula}. The inadequacy of metadata maintenance is evident, as demonstrated by the metadata entry rates for approximately 3.6 million datasets in the Google Dataset Search dataset \cite{googledatasetsearch2019}. While data URLs and titles exhibit an almost 100\% entry rate, the \textit{VariablesMeasured} field (variable information) is only 9\% complete, and \textit{measurementTechnique}, which indicates the data acquisition method, is less than 1\% complete. 

Although research has established that metadata quality directly influences search effectiveness \cite{Koesten2017}, the provision of information regarding the origin and context of the data remains insufficient. Furthermore, data managed on particular platforms are acquired and stored by data providers for diverse purposes, and the development of data catalogs, a unified description format for metadata, and the dissemination of metadata schemas are inadequate, rendering systematic data organization and integration challenging. 

Notably, due to the highly specialized and personalized nature of data, there is no comprehensive taxonomy system in place to provide curation guidance. Despite efforts to develop data catalog vocabularies, such as Data Vocabulary Catalog (DCAT) and Schema.org, sporadically updated and are not being used in a comprehensive manner. Enforcing a uniform format for all the data generated over time and submitted to the platform presents significant difficulties. These challenges exist across all data repository platforms, and impede users from efficiently identifying suitable datasets for data analysis.  This results in critical delays in investment decisions and risk management. Moreover, in scholarly research, the precision and reliability of studies may be compromised by inadequate data or inappropriate data selection. 

The capacity to accurately explore requisite datasets and evaluate their reliability would facilitate the provision of precise data to data users. Consequently, there is a pressing need for a methodology to identify relevant datasets, using whatever limited information is available as metadata as indicators. Based on the aforementioned discussions, the research question of our study is as follows: \textit{How can the semantic similarity among heterogeneous data be evaluated, and how can the accuracy of data discovery be enhanced while addressing insufficient metadata?} This research presents an architecture for data exploration that incorporates Retrieval-Augmented Generation (RAG) into traditional metadata-based evaluations of data similarity, context, and content, yielding results that align with user requirements. RAG enhances the text generation capabilities of Large-Language Models (LLMs) by integrating information from external knowledge bases, potentially mitigating hallucinations associated with standalone LLM use \cite{rag2020}. The proposed framework facilitates the identification of semantic relationships between diverse datasets, surpassing conventional string matching or basic metadata approaches. Moreover, when metadata is insufficient, the architecture can utilize domain knowledge from heterogeneous datasets as supplementary external information to address gaps in the data.

The contributions of this study can be summarized as follows.
\begin{enumerate}
    \item This study represents the initial endeavor to quantitatively assess data similarity utilizing LLM and RAG for heterogeneous metadata retrieval.
    \item This study assessed the tasks of dataset recommendation and estimating data characteristics by diving metadata into textual information (description of the data outline) and structure (variables) as its constituent elements.
    \item Our findings indicate that the optimal model and input method varied according to the type of estimation target and its features, underscoring the significance of selecting an appropriate model based on the specific characteristics of the task.
\end{enumerate}

The subsequent sections of this paper are structured as follows. Section II presents related studies that use our methodological approach. In Section III, we delineate the system architecture and the data models employed for data exploration. Section IV presents the experimental details. In Section V, we show the results and engage in a critical discussion of the study’s limitations. Finally, conclusions are presented in Section VI.

\section{Related Studies}
The analysis of dataset relationships can be classified as three primary approaches: 1) studies focusing on metadata, 2) use of actual data, and 3) external information to depict data connections. Metadata comprises summarized details about a dataset, including its name, summary, tags, and related attributes. The primary content of a dataset, referred to as actual data, is available in various formats such as CSV, JSON, images, texts, and other file types. Knowledge graphs that illustrate data relationships constitute external information.

Among the three approaches, there is general consensus that the one using metadata is the most cost-effective and accurate method for understanding the relationships between different datasets. Metadata typically exploits a unified description framework that is independent of the modality and domain of the dataset. In contrast to approaches based on the contents of actual datasets, metadata facilitates the evaluation of similarity between datasets with differing data structures \cite{hayashi2020understanding, sakaji2021}. An additional advantage of using metadata is that metadata description items are typicall written in natural language, thus enabling the application of numerous NLP methods. Research using metadata-based similarity measures for datasets encompasses vector similarity between word embeddings \cite{Bernhauer2022, Zhang2019, Sakaji2020} and graph distance between ontology concepts \cite{Wang2020, Wang2021, Koda2019}. This study will implement the comparison tasks, such as a similarity evaluation task, utilizing metadata, in accordance with the approach of previous methods.

Furthermore, embedding methods applied to actual datasets, particularly tabular data in instances where adequate metadata is not available, have been proposed in the field of data exploration. For instance, pre-trained models such as Table2Vec \cite{Zhang2019}, TABBIE \cite{TABBIE}, or TAPAS \cite{tapas}, which are capable of converting tabular data into vector representations, have been introduced. Additionally, similarity evaluation using metadata descriptions and variables as data semantic expression extraction focusing on metadata \cite{sakaji2021, hayashi2023} have been developed. However, unlike web articles, actual datasets contain limited textual information. To address this limitation, research has been proposed to employ large language models (LLMs) to facilitate data discovery and semantic understanding using the minimal information available \cite{Dong2024}. Notably, the knowledge and tools required to maximize the potential of LLMs for dataset search have led to substantial advancements in data discovery and semantic understanding \cite{fujita2023, nishio2024}.

As previously discussed, LLMs demonstrate significant potential for data exploration tasks, particularly when employing a metadata-based approach, due to their inherent language-processing capabilities. In this research, we propose a methodology to enhance LLM performance by prioritizing metadata analysis over raw data examination, and to augment domain-specific knowledge through the implementation of the RAG architecture.

\section{System Architecture and Data Modeling}
\subsection{Overview of the Data Exploration System}
\begin{figure}[t]
    \centerline{\includegraphics[width=0.495\textwidth]{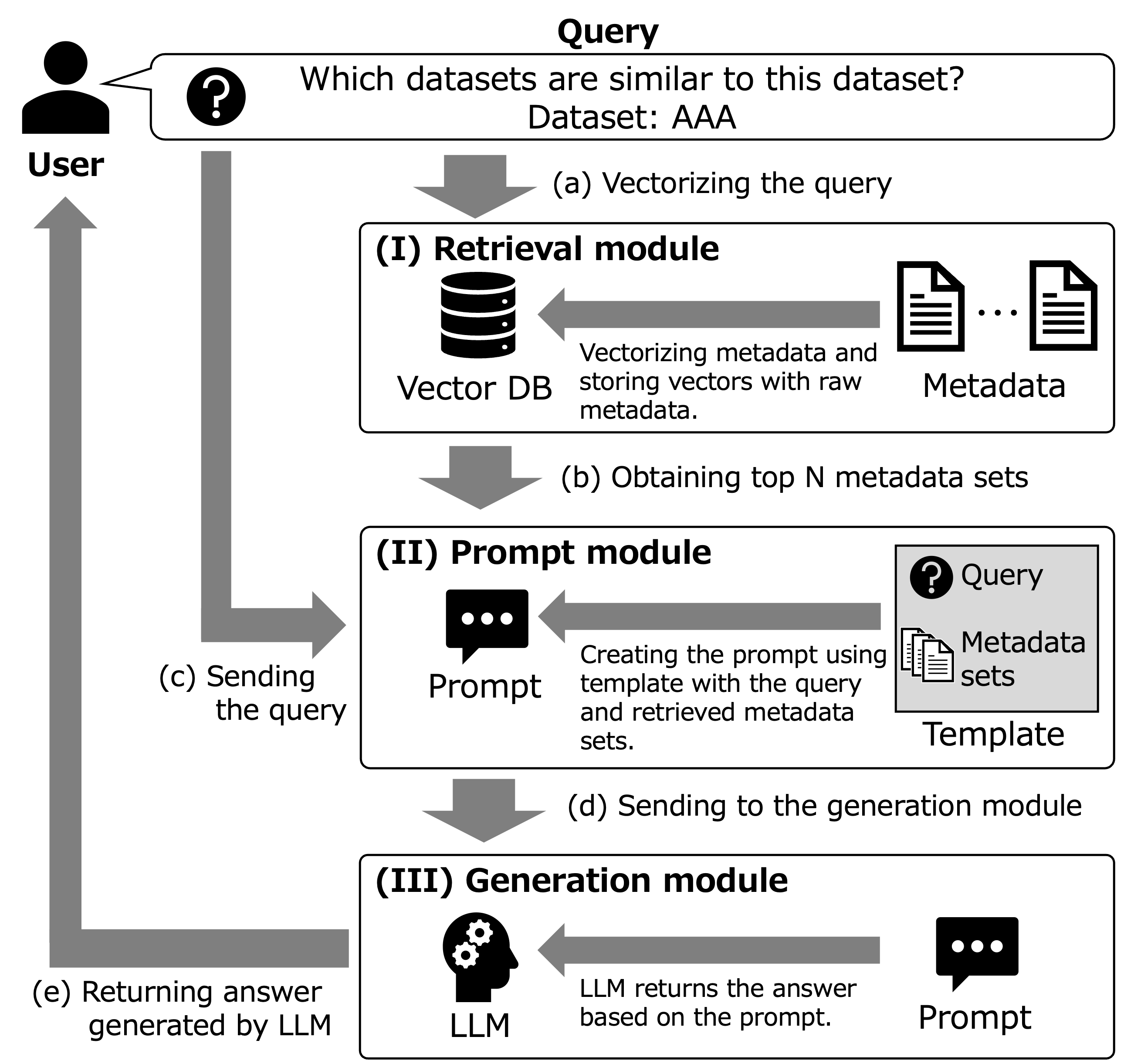}}
    \caption{The proposed RAG system architecture for data exploration.}
    \label{fig:rag-architecture}
\end{figure}

Figure \ref{fig:rag-architecture} illustrates the comprehensive architecture of our RAG system, adjusted and adapted as the data exploration system proposed in this study. Our RAG architecture integrates an LLM with an external database to generate appropriate information for user queries. Initially, in the retrieval module depicted in Figure \ref{fig:rag-architecture}) component labelled ``(I),'' a vector DB is established in advance as an external information source, comprising a vector of embedded metadata representations and the original metadata. The vectorization methodology is elucidated below in Section III-B. To facilitate dataset exploration, the system employs four sequential steps as follows:

\textit{\textbf{Step 1 (Query input and vectorization)}}: To explore a dataset, a user enters a query to obtain information about a dataset that aligns with their interests. The query is formulated in natural language format, where “AAA” represents the name of the dataset. Subsequently, the vectorized query is generated using the same language model employed to create the embedded representation of a vector DB in process (a).

\textit{\textbf{Step 2 (Related dataset search)}}: The user’s vectorized query searches for related datasets from the vector DB and selects candidates. In this step, the top N metadata with high cosine similarity between vectors are obtained in (b).

\textit{\textbf{Step 3 (Prompt creation)}}: The dataset information retrieved from the Retrieval Module is subsequently incorporated into the Prompt module. At this juncture, the user’s query from Step 1 is used again. In this step, the query and associated metadata sets are inserted into a template (details are explained in Section IV) and transmitted to the Generation module at (d).

\textit{\textbf{Step 4 (Answer generation)}}: The prompts are entered into the LLM to generate responses and subsequently return them to the user in (e).

\subsection{Data Modeling by Metadata}
The vector DB, which constitutes the foundation of RAG, is constructed from a set of embedded representations of information on datasets. Given the diverse formats in which data is represented, there is no universally applicable method to describe different types of datasets in a common format. As discussed in Section II, although embedded representations of datasets exist, they are specialized for tabular and graphical data. Therefore, this study calculates the similarity of data pairs based on the elements described in the metadata, which is the data about the data. Our data modeling by metadata is separated into two components: 1) the variable component, which is highly machine-readable, and 2) the description component, which is highly human-readable. Variables represent the attributes of the object that the data represent. Data names and summary descriptions are sentences that provide information about the data in natural language. Table I presents an example of the metadata for the Humanitarian Data Exchange (HDX) cite{https://data.humdata.org/} data utilized in this experiment.

This study employs four language models for metadata items: BERT (bert-base-uncased), Sentence-BERT (intfloat/multilingual-e5-base), Word2Vec (noun), and OpenAI (text-embedding-3-small) to generate embedding representations (details of the models are provided in Section IV-B). This approach enables the computation of similarity between query vectors and metadata. Furthermore, the similarity of dataset pairs can be assessed using cosine similarity. For comparative purposes, this study also utilizes data similarity by variables using the Dice coefficient, referencing the conventional method instead of the embedded representation \cite{Sakaji2020, Sakumoto2024}. The similarity calculation for a dataset pair using the Dice coefficient is as follows:

\begin{equation}
\label{dice_coefficient}
Dice(V_i,V_j) = \frac{2| V_i \cap V_j |}{|V_i|+|V_j|},
\end{equation}

\noindent
where $V_i$ and $V_j$ are the variables of the corresponding datasets $i$ and $j$.

\begin{table}[t]
    \renewcommand{\arraystretch}{1.5}
    \centering
    \caption{Example of HDX Metadata.}\label{tab:metadata}
    \begin{tabular}{l|p{2.5in}}
    \toprule
    \textbf{Item} & \textbf{Content}\\
    \hline
        Data name & Daily Summaries of Precipitation Indicators for Canada  \\
    \hline
        Data summary & This dataset contains the daily summaries on base stations across Canada. The four indicators included are: TPCP: Total precipitation MXSD: Maximum snow depth TSNW: Total snow fall EMXP: Extreme maximum daily precipitation Indicators are compiled by the National Centers for Environmental Information (NCEI), which is administrated by National oceanic and Atmospheric Administration (NoAA) an organization part of the United States government. NoAA has access to data collected from thousands of base stations around the world, which collect data periodically on weather and climate conditions. This dataset contains the latest 5 years of available data.  \\
        \hline
        Variables & 'indicator', 'value', 'station', 'fl\_cmiss', 'date', 'fl\_miss', 'datatype', 'country'  \\
        \hline
        Tags & 'el nino', 'rainfall - precipitation', 'weather and climate'  \\
    \bottomrule
    \end{tabular}
\end{table}

\section{Experimental Settings}
\subsection{Purpose}
Our objective is to assess the efficacy of a customized RAG system for dataset exploration in data retrieval tasks metadata as a constrained source of information about data. Specifically, the evaluation will be conducted employing the following two approaches.

\subsubsection{Differential comparison of data similarity through description-based and variable-based similarity}
Four major language models are used to analyze relationships between data on a textual basis. Additionally, relationships between data will be analyzed on a variable basis. The data similarity across language models will be compared, and the differences and similarities between the models will be examined.

\subsubsection{Consideration of the possibility that different language models should be used for different tasks}
This study compares the efficacy of metadata items and language models across four distinct tasks: similar dataset recommendation, combinable dataset recommendation, tag estimation, and variable estimation. Furthermore, it examines the significance of task-specific selection of metadata items and language models.

\subsection{Language Models and Vector DB}\label{LMs}
Four language models are utilized in the creation of the vector DBs. The first model is BERT, a language model that has demonstrated exceptional performance across various language interpretation tasks \cite{bert}. BERT is pre-trained on BookCorpus \cite{Zhu_2015_ICCV} and Wikipedia entries; if the number of tokens exceeds 512, subsequent tokens are truncated. The second model is Sentence-BERT (hereafter, SBERT) \cite{sbert}. It vectorizes sentences, with the tokenizing functionality integrated into SBERT. The fundamental operation is analogous to vectorization in BERT; however, a model is employed in which the vector output is optimized for similar sentence determination. The vector length is 768 dimensions, and if it exceeds 512 tokens, the remainder are truncated. The third, OpenAI model, functions similarly to SBERT and obtains embedded representations from sentences. The internal model is not publicly accessible. The vector comprises 1,536 dimensions, and if it exceeds 8,191 tokens, the remainder are truncated. The fourth, Word2Vec (hereafter, W2V), is a method that represents words as vectors \cite{mikolov}. In this experiment, only nouns are considered, and it derived from Reuters news articles spanning 2003 to 2018 is utilized. The vectors have 200 dimensions, generated with a 5-word window, and no token truncation is applied. The training algorithm employed was Continuous Bag of Words (CBOW).
% Add by sakaji
In the word2vec vector, we calculate the average of word vectors for getting a sentence vector.

The following three categories of metadata elements were utilized as input to generate embeddings from the language models.

\begin{itemize}
    \item Description model (D): Data name, Summary, and Tags
    \item Variable model (V): Variables and Tags
    \item Description and Variable model (D+V): Data name, Summary, Variables, and Tags
\end{itemize}

In addition, we use ChromaDB\footnote{https://docs.trychroma.com/} for building vector DBs, which is efficient for vector search, and Llama 3.1\footnote{https://www.llama.com/} for LLM.

\subsection{Four Tasks and Evaluation Methods}
We implemented four tasks that are considered significant in dataset search. The first is a similar dataset recommendation task (Task 1). This task involves recommending datasets that are similar to a given dataset, and is considered the most fundamental task in dataset search, having been implemented in various studies \cite{Sakumoto2024}. The second is a combinable dataset recommendation task (Task 2). This task reflects the observation that, as demonstrated in \cite{rsoc-hayashi}, multiple data sources are increasingly being combined for analysis, rather than relying on a single data source. The third is tag estimation (Task 3). There exist instances where metadata creators are uncertain about which tags to assign to their own data when providing their datasets. Additionally, metadata creation is considered burdensome. This task predicts the category to which the dataset in question belongs as tags. The fourth task is variable estimation (Task 4), which predicts potential variables within the dataset \cite{variablequest}. Data design generally incurs high rework costs, and it is beneficial to estimate variables from data summaries prior to data acquisition. Each task is presented as a query in natural language sentences as follows:

\textit{Task 1}: “Which datasets are similar to this dataset? Please select multiple candidates and rank them by relevance.”

\textit{Task 2}: “Which datasets would be suitable to combine with this dataset? Please select multiple candidates and rank them by their combinability.”

\textit{Task 3}: “What tags are associated with this dataset? Please select from the list below and sort the results by relevance.”

\textit{Task 4}: “What variables are included in this dataset? Please select from the list below and sort the results by relevance.”

The query is vectorized by the language models and N related metadata are obtained from the vector DBs (N=10 in the experiment). The query, consisting of natural language sentences, and 10 metadata expressions are subsequently inserted into the prompt template to generate the results from LLM.

In Task 1, we assess the correspondence between the tags in the sample datasets (described in detail in Section IV-D) and those in the output datasets from the LLM. If the tags are congruent, the recommendation is classified as originating from the same category; if they are incongruent, the recommendation is categorized as arising from a different category. Subsequently, the similarity of the dataset pair is analyzed in terms of description and variables. As in Task 1, Task 2 is evaluated based on the correspondence between the tags in the sample dataset and those in the output datasets generated by LLM. It is important to note that dataset combinability is not necessarily advantageous even when the dataset pair originates from the same category. Given that the presence or absence of common variables, such as foreign keys, is crucial for data integration and combination, the combinability of datasets is assessed in terms of variable similarity. The variable model (V) generates a collection of variables and tags as metadata from the vector DB. However, the output of V lacks the data name information that uniquely identifies the relevant dataset. Consequently, we compare the outputs between D and D+V in Tasks 1 and 2.

Given that Tasks 3 and 4 are estimation tasks, we evaluate the Precision, Recall, and F1 scores of the output results from the LLM. For the three inputs D, V, and D+V, we analyze the differences in results with respect to the characteristics of each language model.

\subsection{Datasets}
Using the appropriate API, we extracted 9,630 datasets (provided in CSV format) from the HDX website\footnote{https://data.humdata.org/}. These datasets included information such as the data name, summary, variables, and associated tags. Table II presents the fundamental statistical features of these datasets.

\begin{table}[t]
    \renewcommand{\arraystretch}{1.5}
    \centering
    \caption{The Features of Variables and Tags.}\label{tab:features}
    \begin{tabular}{p{.9in}c|p{.9in}c}
    \toprule
        \textbf{Variable} & \textbf{Value} & \textbf{Tag} & \textbf{Value}\\
    \hline
        \# of variables & 149,582 & \# of tags & 42,379\\
    \hline
        \# of variable types & 13,409 & \# of tag types & 249\\
    \hline
        Max \# of variables in a dataset & 2,856 & Max \# of tags in a dataset & 19\\
    \hline
        Min \# of variables in a dataset & 1 & Min \# of tags in a dataset & 1\\
    \bottomrule
    \end{tabular}
\end{table}

To evaluate the performance of the four tasks, a sample dataset was prepared, comprising complete data names, data summaries, variables, and tags. Five tags with high frequency of occurrence—education, economics, health, facilities and infrastructure, and weather and climate—were selected. Two datasets with each tag were extracted to ensure the tags did not overlap, resulting in a total of 10 datasets. Table III presents the sample datasets for validation.

This methodology enables the tracking of discrepancies in input metadata items and similarities among the target data categories by tags. The evaluation of differences in the performance of the language model, similarity calculation method, and task will be feasible based on the type of question and the data categories covered by the sample dataset.

\begin{table}[t]
    \renewcommand{\arraystretch}{1.5}
    \centering
    \caption{Sample Datasets.}\label{tab:metadata}
    \begin{tabular}{p{0.7in}|p{2.5in}}
    \toprule
    \textbf{Category} & \textbf{Data name}\\
    \hline
        Education &
        \begin{itemize}
            \item Compiled Reports of the Special Representative of the Secretary General for Children and Armed conflict of years 2015 through 2017
            \item UNHCR's populations of concern originating from the Democratic Republic of the Congo
        \end{itemize}\\
        \hline
        Economics &
        \begin{itemize}
            \item Japan - Social Development
            \item Canada - Science and Technology
        \end{itemize}\\
        \hline
        Health &
        \begin{itemize}
            \item InterAction member activities in Greece
            \item Airports in Cayman Islands
        \end{itemize}\\
        \hline
        Facilities and infrastructure &
        \begin{itemize}
            \item South Sudan 3W operational Presence (Jan - Apr-2016)
            \item Current IATI aid activities in Guinea-Bissau
        \end{itemize}\\
        \hline
        Weather and climate &
        \begin{itemize}
            \item Japan - Economic, Social, Environmental, Health, Education, Development and Energy
            \item Daily Summaries of Precipitation Indicators for Canada
        \end{itemize}\\
    \bottomrule
    \end{tabular}
\end{table}

\subsection{Prompt Generation}
The method for generating a prompt for nput into the LLM comprises two steps: (1) retrieving the top $N$ datasets (metadata) from the vector DB through a vectorized query, and (2) incorporating the query and the top $N$ datasets into the predetermined template. The template structure is as follows, wherein \{question\} represents the query and \{context\} denotes the list of the top $N$ datasets.

\begin{tcolorbox}
You are an assistant for question-answering tasks. Use the following pieces of retrieved context to answer the question. If you don’t know the answer, just say that you don’t know. Use five sentences maximum and keep the answer concise.

Question: \{question\}

Context: \{context\}
\end{tcolorbox}

\section{Results and Discussion}
\subsection{Evaluation on Similar Dataset Recommendation (Task 1)}
Figure 2 (a) shows the output results of similar dataset recommendation (Task 1) with four language models and two metadata input types. For each sample metadata, the vector DBs return 10 related metadata with queries across all models, totaling 100. The quantity was subsequently reduced by more than half, limited to those deemed most relevant by the LLM. Furthermore, some models incorporate a separate fictional dataset generated by the LLM itself.

Initially, we compare the number of dataset origins by model. For the OpenAI (D), it is observed that a greater number of datasets are generated by LLM compared to other models. Upon closer examination, it became apparent that the majority of the information was inaccurate, indicating the occurrence of hallucination. This false data originated from the health-tagged “InterAction member activities in Greece” dataset, despite the fact that the metadata sets retrieved from the vector DB produced 10 metadata showing high similarity to the query. Therefore, it does not necessarily indicate that the LLM with OpenAI model possesses the capability to generate more datasets than other models.

A shared feature is evident across all four models. Model Ds exhibits a higher number of datasets from different categories compared to those from the same category as the sample datasets. Additionally, when contrasted with D, the quantity of recommendations by LLM decreases with adding variable information in D+V. 

With each LLM, we found that the datasets determined to be more similar were selected from the results by vector DBs. However, was any of the LLMs able to select more similar datasets? Figure 3 compares the mean similarity by Variable and Description before and after conducting selection by LLM. Common to all models is that the datasets recommended from the vector DBs before LLM perform relatively well in recommending datasets from the same category, but not well enough for recommending datasets from different categories. Notably, all models except W2V demonstrated an ability to suggest datasets with greater similarity in both variables and descriptions through an LLM. Of note is BERT (D+V), which successfully identified datasets with higher variable similarity from different categories using LLM. This trend was observed across other models, where LLM facilitated the selection of more similar datasets from categories different from the sample data. Although recommendations within the same category were not as effective as those from different categories, LLMs consistently identified datasets with higher similarity.

In summary, in three models of D and D+V, except W2V, the variable similarity is improved by using an LLM. In particular, it is possible to select datasets with high variable similarity in datasets that belong to a different category from the sample data. The datasets in the same category do not differ greatly, but the average similarity is slightly improved. It may be that LLM may be superior in selecting datasets with high variable similarity that could not be picked up by the vector DB. This result also suggests that the influence of Description is significant in the similar dataset recommendation task. Therefore, in the same category, datasets with similar Descriptions tend to have similar variables, and as a result, datasets with high similarity are more likely to be recommended. On the other hand, in different categories, datasets with high similarity in Description tend to be recommended, but the result is that the variable similarity of such data is not necessarily high.

\begin{figure*}[t]
    \centerline{\includegraphics[width=1.0\textwidth]{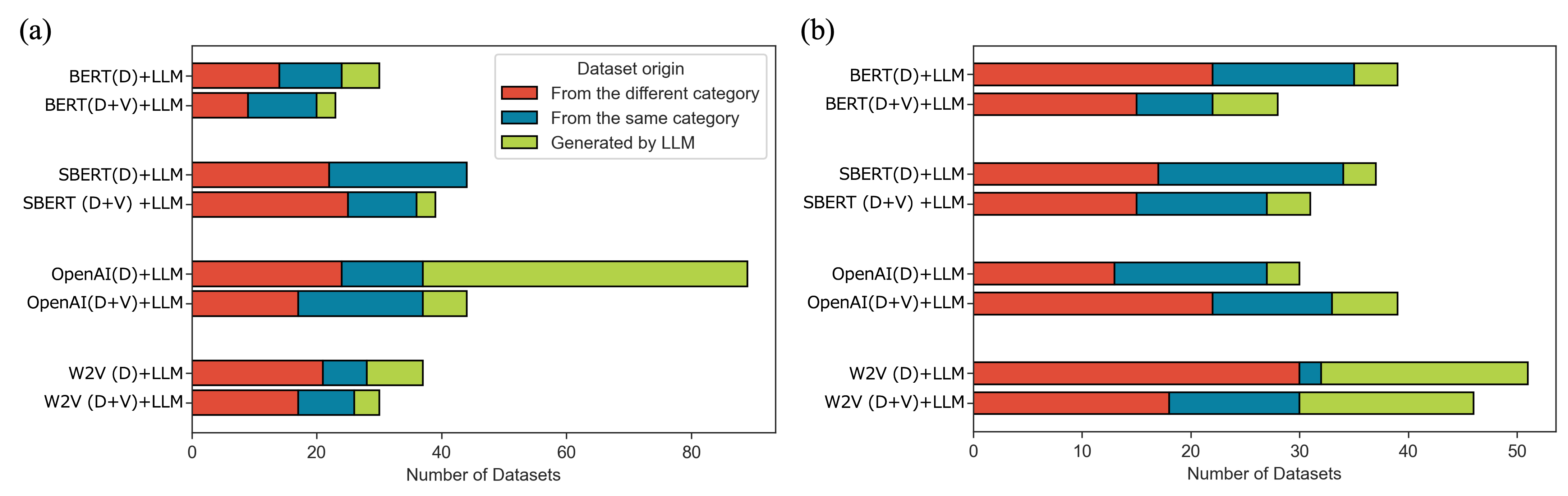}}
    \caption{Output results of (a) similar dataset recommendation (Task 1) with four language models and three metadata input types and (b) combinable dataset recommendation (Task 2). The both bar graphs depict the sources of recommended datasets classified into three groups: “datasets from different categories (red),” “datasets from the same category (blue),” and “datasets generated by LLM (lightgreen).” The first two groups consist of actual datasets in HDX, while the third category comprises fictional datasets generated by LLM.}
    \label{fig:task1_2}
\end{figure*}

\begin{figure*}[t]
    \centerline{\includegraphics[width=1.0\textwidth]{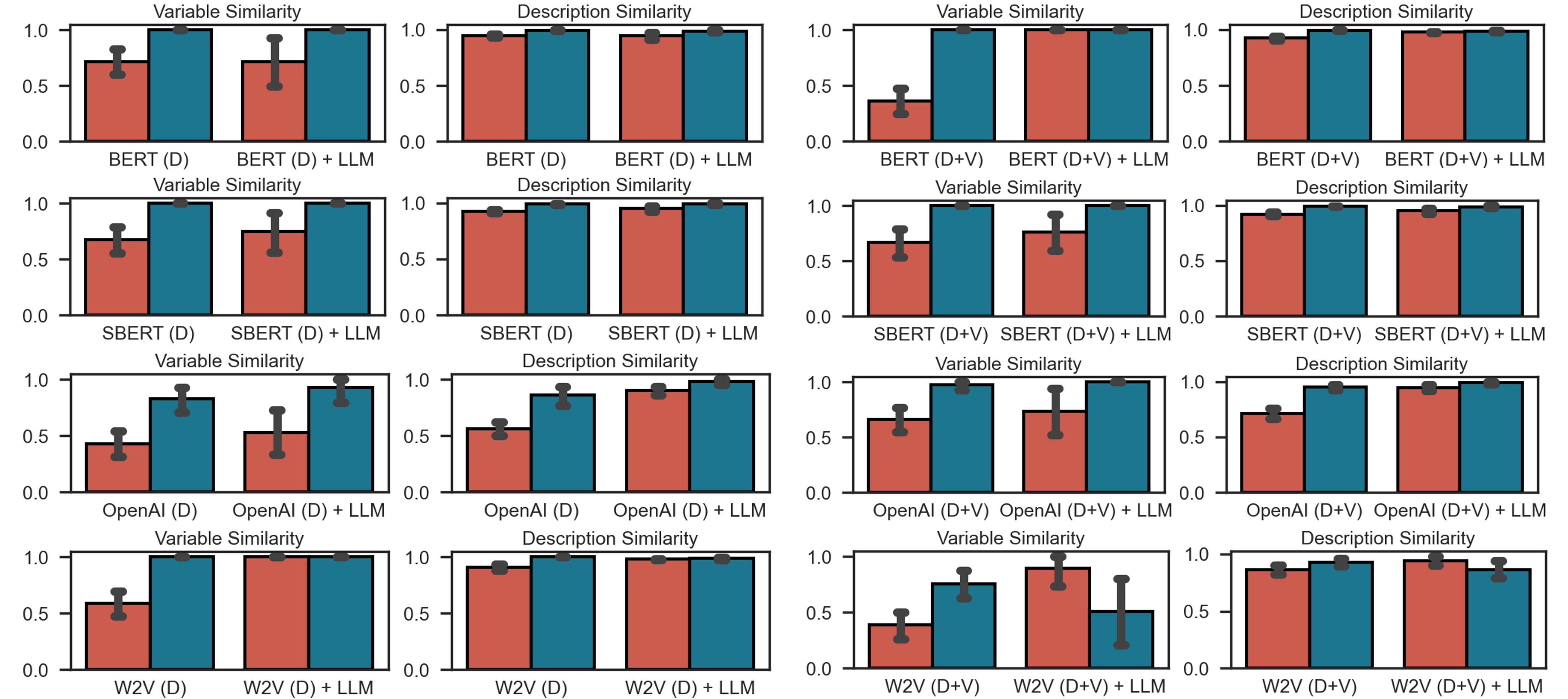}}
    \caption{Comparison of mean variable/description similarities of recommended datasets before and after via LLM in Task 1. The bar graphs depict the sources of recommended datasets classified into two groups: “datasets from different categories (red),” and “datasets from the same category (blue).” The error bars show the standard deviation.}
    \label{fig:task1_result}
\end{figure*}

\subsection{Evaluation of Combinable Dataset Recommendation (Task 2)}
Figure 2 (b) shows the output results of combinable dataset recommendation (Task 2) with four language models and two metadata input types. As with Task 1, the number of outputs has been reduced by more than half using LLM. If the number of LLM-generated datasets of OpenAI (D) in Task 1, in which are hallucination is excluded as noise, then there is an overall increase in the datasets generated by LLM in Task 2 (61) compared to that of Task 1 (32). This is likely due to the greater complexity of the combinable dataset recommendation task and the increased need to generate new data sets by LLM. This is especially true for inputs that include variable information with description in V+D, suggesting that LLM is attempting to make better recommendations by generating new combinations in addition to existing datasets.

Was the LLM able to select more combinable datasets? Figure 4 compares the mean similarity by Variable and Description before and after conducting selection by LLM. Common to all models is that the datasets recommended from the vector DBs before LLM perform relatively well in recommending datasets from the same category, but not well enough for recommending datasets from different categories. Notably, all models except W2V demonstrated an ability to suggest datasets with greater similarity in both variables and descriptions through LLM. Especially, BERT (D+V) successfully identified datasets with higher variable similarity from different categories using LLM. This trend was observed across other models, where LLM facilitated the selection of more similar datasets from categories different from the sample data. Although recommendations within the same category were not as effective as those from different categories, LLMs consistently identified datasets with higher similarity.

How combinable are the recommended datasets with the sample dataset? Specifically, given that the integration of heterogeneous datasets typically occurs through shared variables, as exemplified by foreign keys, variable similarity is considered more crucial than description similarity. In recommendations from the same category using BERT and SBERT, datasets exhibiting high similarity to the sample dataset in both variables and descriptions were recommended from vector DBs. This indicates that datasets with high potential for combination are proposed without the need for LLM application. Conversely, in recommendations from different categories, the similarity of variables significantly improves through LLM in BERT (D+V), SBERT (D), and SBERT (D+V). This demonstrates that LLM selected datasets with more comparable variables. A similar trend is observed for OpenAI and W2V. These findings suggest that LLM may exhibit superior performance in identifying datasets that are more likely to be combined when recommending data from a different domain to be merged with the sample datasets.

While LLM enhances the correlation between variables and descriptions in most of the models, it notably {\it decreases} variable similarity within the same categories for W2V (D+V). This suggests that LLM might exclude certain combinable datasets. In Task 1, the W2V model underperformed compared to other models in D+V. Although W2V outperforms OpenAI in D, its performance declines when variable information is incorporated in D+V. This decline may be attributed to the noun-centric nature of W2V, which is highly customized for human understanding. Variable information, being more specialized for machine interpretation than human comprehension, likely becomes noise in the W2V model during the embedding process.

\begin{figure*}[t]
    \centerline{\includegraphics[width=1.0\textwidth]{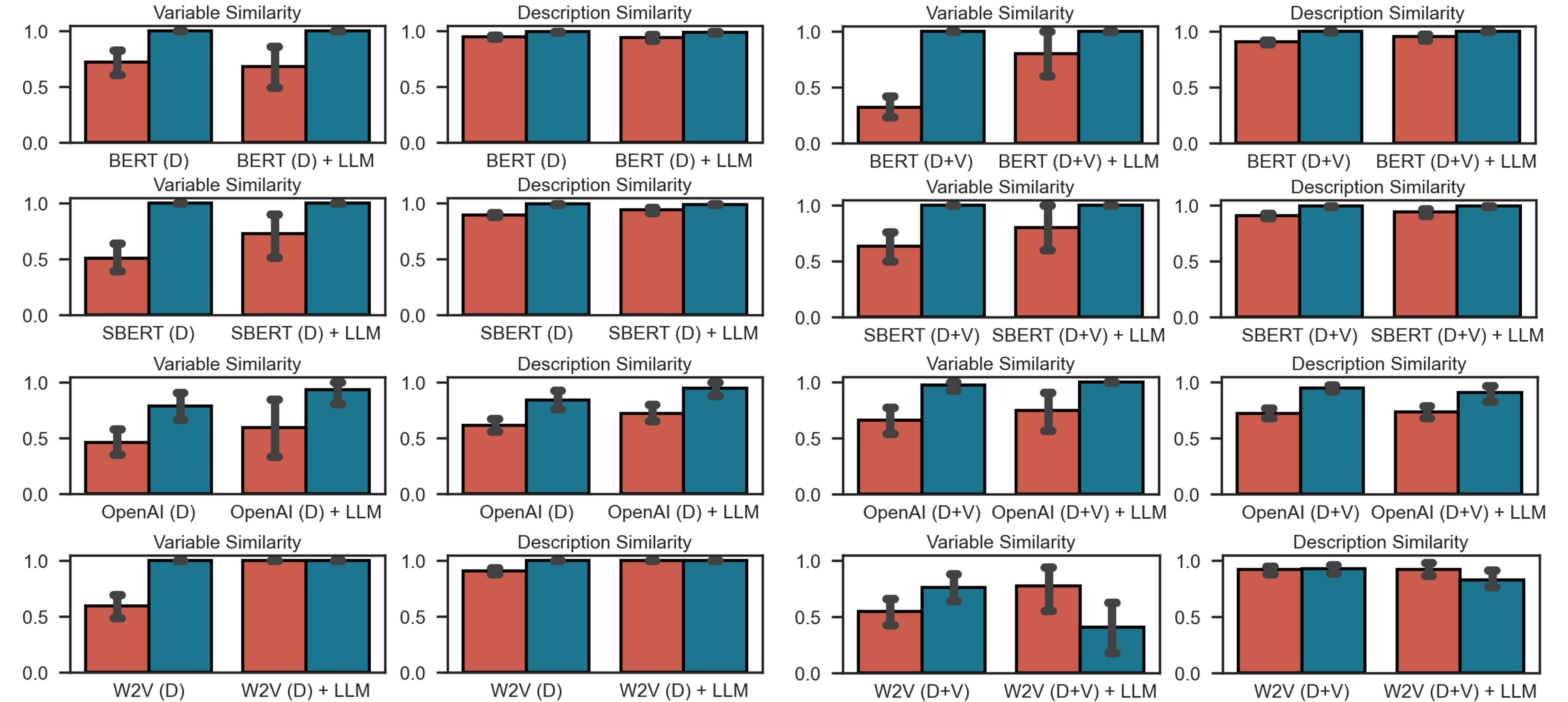}}
    \caption{Comparison of mean variable/description similarities of recommended datasets before and after via LLM in Task 2. The bar graphs depict the sources of recommended datasets classified into two groups: “datasets from different categories (red),” and “datasets from the same category (blues).” The error bars show the standard deviation.}
    \label{fig:task2_result}
\end{figure*}

\begin{figure*}[t]
    \centerline{\includegraphics[width=1.0\textwidth]{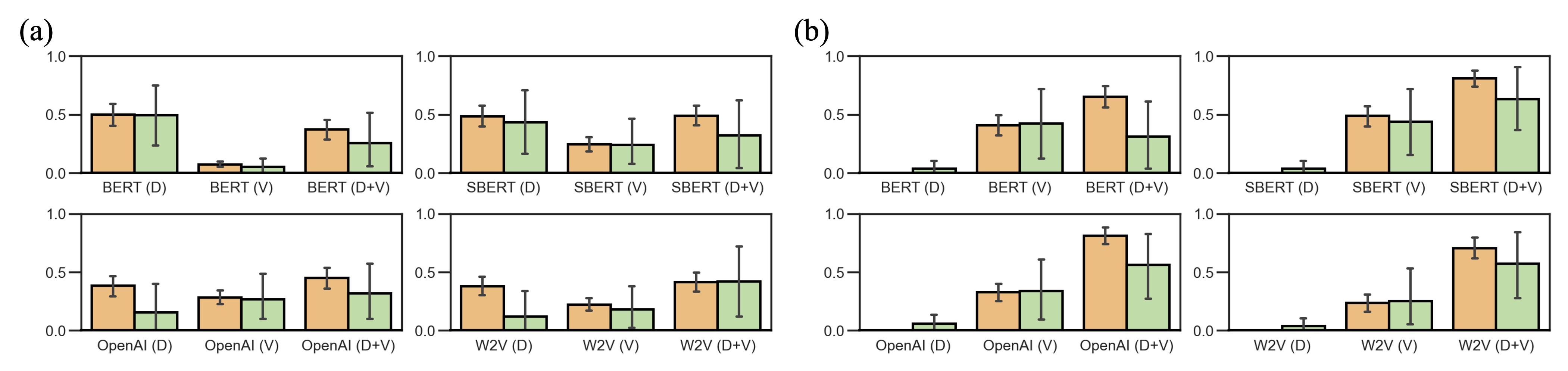}}
    \caption{Mean F1 scores in (a) Task 3, and (b) Task 4. The bar graphs on the left of each model (orange) are the scores for the tags/variables contained in the metadata output from the vector DBs, and the ones on the right (green) are the scores for those selected via LLM.}
    \label{fig:task2_result}
\end{figure*}

\subsection{Evaluation of Tag Estimation (Task 3)}
The average F1 Scores for Task 3 are depicted in Fig. 5 (a). Models that us D alone or D+V demonstrated high performance, whereas those employing V embeddings exhibited extremely low performance. This discrepancy arises because tag information is contingent on the context and topic of the sample dataset, resulting in superior performance for text-based inputs (D and D+V). The incorporation of V in D+V models did not yield significant improvements compared to D alone, and in certain instances, scores actually declined. This suggests that variable information fails to serve as effective supplementary information for tag estimation. It appears that the variable information lacks sufficient determining characteristics representative of the data categories denoted by the tags, and instead functions as noise in the process. For tag estimation, BERT and SBERT’s D alone without LLM performs best. This result suggests that these models can most effectively use natural language text as input; BERT and SBERT were trained on a large amount of text data during pre-training, and thus have very high comprehension of the text with D. 

The introduction of LLM resulted in decreased scores or increased standard deviations across all models, suggesting performance instability. Examination of individual outputs revealed that the LLM excluded correct tags identified as relevant in the vector DBs. Furthermore, the LLM generated its own tags not present in the HDX tag list. The current LLM’s use in tag estimation tasks leads to unreliable results, indicating that its application should be carefully evaluated based on the specific task requirements.

\subsection{Evaluation of Variable Estimation (Task 4)}
Similar to Task 3, the output from Task 4 utilizing vector DB alone consistently yielded superior F1 scores across all models compared to using LLM (Fig. 5 (b)). An analysis of individual output results revealed that accurate variables identified by vector DBs were omitted during the LLM process, or the LLM generated distinct variables not present in the HDX variable sets. The application of current LLM for variable estimation tasks should be approached with caution.

Interestingly, the combination of description and variable information (D+V) outperforms variable information alone (V) across all models. This suggests that, contrary to tag information, variable data is not extraneous in variable estimation, and its integration with descriptive content enables more accurate estimation of relevant variable sets. For instance, in the “Japan - Social Development” dataset, the correct variables are “value,” “indicator name,” “country iso3,” “year,” “indicator code,” and “country name.” The SBERT (V) model estimated a variable set including “value,” “origin,” “year,” and “country / territory of asylum/residence,” achieving partial correctness with an F1 score of 0.40. This partial match occurred because the query's sample dataset description partially aligned with the variable information in the vector DB. In contrast, when using the complete metadata (description and variables), using the D+V model resulted in a higher similarity output for the dataset and its variables compared to using variables alone, yielding a perfect F1 score of 1.00. While the variable-only model demonstrates some estimation capability, incorporating descriptive information, which provides context for the variables, appears to have supplemented the variable data, leading to enhanced estimation performance.

In summary, we found that enhancing variable estimation with description information is more effective than relying solely on variable information. Notably, despite variable estimation being more challenging than tag estimation, it generally achieves higher scores. This is unexpected, given that variable estimation involves 13,409 variables compared to 249 tags, and variables exhibit a longer-tailed frequency distribution. The superior performance of variable estimation may be attributed to the co-occurrence nature of variables. For instance, while “country / territory of asylum/residence” has moderate frequency, it frequently appears alongside high-frequency variables like “year” and “origin,” resulting in a high number of variable pair occurrences. This co-occurrence pattern, where low-frequency variables sometimes appear with high-frequency ones, is believed to boost variable estimation performance. Conversely, tags demonstrate different characteristics. As shown in Table II, datasets contain a maximum of 19 tags, with an average of 4.4 tags per dataset, resembling a Poisson distribution. Unlike variables, tags are independently assigned as representative topics for datasets, rather than co-occurring. This independence is thought to limit tag estimation performance.

\section{Conclusion}
The proposed RAG-based system demonstrated good performance in identifying similar and combinable datasets. Additionally, LLMs displayed proficiency in choosing datasets with greater variable and description similarities, particularly from diverse categories, surpassing the capabilities of vector DB retrieval alone. However, performance fluctuated across different tasks and models, emphasizing the necessity of carefully selecting appropriate techniques and metadata items based on specific use cases, and indicating that caution should be exercised when applying LLMs to these tasks. In essence, while LLMs enhanced results for similar and combinable dataset recommendations, vector DB retrieval might be more suitable for tag and variable estimation tasks. 

Although this approach shows potential for tackling challenges in data exploration and discovery, further improvements are needed, especially for estimation tasks where LLMs introduced inconsistencies. 

To conclude, this research illustrates the potential of RAG-based systems to improve data exploration capabilities, particularly in identifying relationships between heterogeneous datasets. Nevertheless, the inconsistent performance across tasks underlines the importance of carefully considering model and input selection based on specific data discovery goals. For example, instead of relying on generic language models, more domain-specific models could be considered for better capturing the semantics of particular database inputs. In addition, deploying explainable AI methods that indicate what database features most influence similarity measures could provide insights on the knowledge captured by language models and the decision-making of the overall database retrieval architecture. Future research should concentrate on enhancing LLM applications for estimation tasks and investigating methods to more effectively combine the strengths of both vector DB retrieval and language models.

\section*{Acknowledgment}
This study was supported by JST PRESTO Grant Number JPMJPR2369.

\bibliographystyle{IEEEtran}
\bibliography{ref}

\begin{thebibliography}{10}
\providecommand{\url}[1]{#1}
\csname url@samestyle\endcsname
\providecommand{\newblock}{\relax}
\providecommand{\bibinfo}[2]{#2}
\providecommand{\BIBentrySTDinterwordspacing}{\spaceskip=0pt\relax}
\providecommand{\BIBentryALTinterwordstretchfactor}{4}
\providecommand{\BIBentryALTinterwordspacing}{\spaceskip=\fontdimen2\font plus
\BIBentryALTinterwordstretchfactor\fontdimen3\font minus \fontdimen4\font\relax}
\providecommand{\BIBforeignlanguage}[2]{{%
\expandafter\ifx\csname l@#1\endcsname\relax
\typeout{** WARNING: IEEEtran.bst: No hyphenation pattern has been}%
\typeout{** loaded for the language `#1'. Using the pattern for}%
\typeout{** the default language instead.}%
\else
\language=\csname l@#1\endcsname
\fi
#2}}
\providecommand{\BIBdecl}{\relax}
\BIBdecl

\bibitem{chapman2020}
A.~Chapman, E.~Simperl, L.~Koesten, G.~Konstantinidis, L.-D. Ibáñez, E.~Kacprzak, and P.~Groth, ``Dataset search: a survey,'' \emph{The VLDB Journal}, vol.~29, pp. 251--272, 2020.

\bibitem{miller2018opendatatransparent}
R.~J. Miller, F.~Nargesian, E.~Zhu, C.~Christodoulakis, K.~Q. Pu, and P.~Andritsos, ``Making open data transparent: Data discovery on open data,'' \emph{{IEEE} Data Engineering Bulletin}, vol.~41, no.~2, pp. 59--70, 2018.

\bibitem{zezula}
P.~Zezula, ``Similarity searching for the big data,'' \emph{Mobile Networks and Applications}, vol.~20, no.~4, p. 487–496, aug 2015.

\bibitem{googledatasetsearch2019}
D.~Brickley, M.~Burgess, and N.~Noy, ``Google dataset search: Building a search engine for datasets in an open web ecosystem,'' \emph{The World Wide Web Conference}, p. 1365–1375, 2019.

\bibitem{Koesten2017}
L.~M. Koesten, E.~Kacprzak, J.~F.~A. Tennison, and E.~Simperl, ``The trials and tribulations of working with structured data: -a study on information seeking behaviour,'' \emph{CHI}, 2017.

\bibitem{rag2020}
P.~Lewis, E.~Perez, A.~Piktus, F.~Petroni, V.~Karpukhin, N.~Goyal, H.~Küttler, M.~Lewis, W.~tau Yih, T.~Rocktäschel, S.~Riedel, and D.~Kiela, ``Retrieval-augmented generation for knowledge-intensive nlp tasks,'' \emph{NeurIPS}, 2020.

\bibitem{hayashi2020understanding}
T.~Hayashi and Y.~Ohsawa, ``Understanding the structural characteristics of data platforms using metadata and a network approach,'' \emph{IEEE Access}, vol.~8, pp. 35\,469--35\,481, 2020.

\bibitem{sakaji2021}
H.~Sakaji, T.~Hayashi, Y.~Fukami, T.~Shimizu, H.~Matsushima, and K.~Izumi, ``Retrieving of data similarity using metadata on a data analysis competition platform,'' \emph{International Conference on Big Data}, pp. 3480--3485, 2021.

\bibitem{Bernhauer2022}
D.~Bernhauer, M.~Nečaský, P.~Škoda, J.~Klímek, and T.~Skopal, ``Open dataset discovery using context-enhanced similarity search,'' \emph{Knowledge and Information Systems}, vol.~64, p. 3265–3291, 2022.

\bibitem{Zhang2019}
L.~Zhang, S.~Zhang, and K.~Balog, ``Table2vec: Neural word and entity embeddings for table population and retrieval,'' \emph{SIGIR}, p. 1029–1032, 2019.

\bibitem{Sakaji2020}
H.~Sakaji, T.~Hayashi, K.~Izumi, and Y.~Ohsawa, ``Verification of data similarity using metadata on a data exchange platform,'' \emph{International Conference on Big Data}, pp. 4467--4474, 2020.

\bibitem{Wang2020}
X.~Wang, Z.~Huang, and F.~van Harmelen, ``Evaluating similarity measures for dataset search,'' \emph{International Conference on Web Information Systems Engineering}, 2020.

\bibitem{Wang2021}
X.~Wang, F.~van Harmelen, and Z.~Huang, ``Biomedical dataset recommendation,'' \emph{10th International Conference on Data Science, Technology and Applications}, vol.~1, pp. 192--199, 2021.

\bibitem{Koda2019}
P.~Škoda, J.~Klímek, M.~Nečaský, and T.~Skopal, ``Explainable similarity of datasets using knowledge graph,'' \emph{International Conference on Similarity Search and Applications}, p. 103–110, 2019.

\bibitem{TABBIE}
H.~Iida, D.~Thai, V.~Manjunatha, and M.~Iyyer, ``{TABBIE}: Pretrained representations of tabular data,'' \emph{Conference of the North American Chapter of the Association for Computational Linguistics}, pp. 3446--3456, 2021.

\bibitem{tapas}
J.~Herzig, P.~K. Nowak, T.~M{\"u}ller, F.~Piccinno, and J.~M. Eisenschlos, ``Tapas: Weakly supervised table parsing via pre-training,'' \emph{58th Annual Meeting of the Association for Computational Linguistics}, 2020.

\bibitem{hayashi2023}
T.~Hayashi, Y.~Fujita, and M.~Kuwahara, ``Exploring the fundamental units of semantic representation of data using heterogeneous variable network in data ecosystems,'' \emph{International Conference on Big Data}, 2023.

\bibitem{Dong2024}
H.~Dong and Z.~Wang, ``Large language models for tabular data: Progresses and future directions,'' \emph{SIGIR}, 2024.

\bibitem{fujita2023}
Y.~Fujita, T.~Hayashi, and M.~Kuwahara, ``Topic-based search: Dataset search without metadata and users’ knowledge about data,'' 2023.

\bibitem{nishio2024}
S.~Nishio, H.~Nonaka, N.~Tsuchiya, A.~Migita, Y.~Banno, T.~Hayashi, H.~Sakaji, T.~Sakumoto, and K.~Watabe, ``Extraction of research objectives, machine learning model names, and dataset names from academic papers and analysis of their interrelationships using llm and network analysis,'' \emph{https://arxiv.org/abs/2408.12097}, 2024.

\bibitem{Sakumoto2024}
T.~Sakumoto, T.~Hayashi, H.~Sakaji, and H.~Nonaka, ``Metadata-based clustering and selection of metadata items for similar dataset discovery and data combination tasks,'' \emph{IEEE Access}, vol.~12, pp. 40\,213--40\,224, 2024.

\bibitem{bert}
J.~Devlin, M.-W. Chang, K.~Lee, and K.~Toutanova, ``Bert: Pre-training of deep bidirectional transformers for language understanding,'' \emph{NAACL}, p. 4171–4186, 2019.

\bibitem{Zhu_2015_ICCV}
Y.~Zhu, R.~Kiros, R.~Zemel, R.~Salakhutdinov, R.~Urtasun, A.~Torralba, and S.~Fidler, ``Aligning books and movies: Towards story-like visual explanations by watching movies and reading books,'' in \emph{The IEEE International Conference on Computer Vision (ICCV)}, December 2015.

\bibitem{sbert}
N.~Reimers and I.~Gurevych, ``Sentence-bert: Sentence embeddings using siamese bert-networks,'' \emph{EMNLP}, 2019.

\bibitem{mikolov}
T.~Mikolov, K.~Chen, G.~S. Corrado, and J.~Dean, ``Efficient estimation of word representations in vector space,'' \emph{International Conference on Learning Representations}, 2013.

\bibitem{rsoc-hayashi}
T.~Hayashi, H.~Sakaji, Y.~F. Hiroyasu~Matsushima, T.~Shimizu, and Y.~Ohsawa, ``Data combination for problem-solving: A case of an open data exchange platform,'' \emph{The Review of Socionetwork Strategies}, vol.~15, p. 521–534, 2021.

\bibitem{variablequest}
T.~Hayashi and Y.~Ohsawa, ``Matrix-based method for inferring variable labels using outlines of data in data jackets,'' \emph{PAKDD}, pp. 696--707, 2017.

\end{thebibliography}

\end{document}